\documentclass[12pt]{article}

\usepackage{epsfig}

\usepackage{amsmath}
\usepackage{amssymb}
\usepackage{euscript}

\newcommand{\Peu}{\EuScript{P}}

\begin{document}

\begin{titlepage}


\begin{flushright}
\bf HNINP-V-04-01
\end{flushright}

\vspace{1mm}
\begin{center}
{\LARGE\bf
  Exact solutions \\
  of the QCD evolution equations \\
  using Monte Carlo method$^{\star}$
}
\end{center}
\vspace{3mm}

\begin{center}
{\bf S. Jadach}
{\em and}
{\bf M. Skrzypek} \\

\vspace{1mm}
{\em Institute of Nuclear Physics HNINP-PAS,\\
  ul. Radzikowskiego 152, 31-342 Cracow, Poland,}\\
\end{center}

\vspace{15mm}
\begin{abstract}
We present the exact and precise ($\sim$0.1\%) numerical solution
of the QCD evolution equations 
for the parton distributions
in a wide range of $Q$ and $x$ using Monte Carlo (MC) method,
which relies on the so-called Markovian algorithm.
We point out certain advantages of such a method
with respect to the existing non-MC methods.
We also formulate a challenge of constructing non-Markovian
MC algorithm for the evolution equations for the initial state
QCD radiation with tagging
the type and $x$ of the exiting parton.
This seems to be within the reach of
the presently available computer CPUs
and the sophistication of the MC techniques.
\end{abstract}

\vspace{10mm}
\begin{center}
\em To be submitted to Acta Physica Polonica
\end{center}

\vspace{15mm}
\begin{flushleft}
{\bf HNINP-V-04-01\\
     December~2003}
\end{flushleft}

\vspace{5mm}
\footnoterule
\noindent
{\footnotesize
$^{\star}$Work partly supported by
Polish Government grants
KBN 5P03B09320,   
2P03B00122        
and
the European Community's Human Potential Programme 
``Physics at Colliders'' under the contract HPRN-CT-2000-00149.
}

\end{titlepage}

\section{Introduction}
It is commonly known that the so called evolution equations of 
the quark and gluon distributions in the hadron,
derived in QED and QCD using the renormalization group or diagrammatic
techniques~\cite{DGLAP},
can be interpreted probabilistically as a Markovian process.
This process consists of the random steps forward in the logarithm
of the energy scale, $t=\ln Q$, and the collinear decay of partons.
Such a Markovian process can be readily used as a basis of the
Monte Carlo (MC) algorithm producing events in the so called
Parton Shower MC (PSMC), see for example ref.~\cite{webber88}.
The Markovian MC provides, in principle, an exact solution
of the evolution equations.
This possibility was not exploited in practice in the past,
mainly because alternative non-MC numerical methods and programs
solving the QCD evolution equations on the finite grid of points
in the space of $t$ and parton energy $x$
are much faster than the MC method.
Typical non-MC program of this type is {\tt QCDnum16}
of ref.~\cite{qcdnum16}. 
For comparisons with other codes see ref.~\cite{blumleinn96}.

In this work we test practical feasibility of applying 
the MC Markovian method to solve
the QCD evolution equations exactly and precisely (per-mill level),
exploiting computer CPU power available today,
in a wide range of $\ln Q$ and $x$.

Apart from pure exploratory aspect, we believe that MC
method has certain advantages over non-MC methods,
see discussion below.

It is also commonly known that the basic Markovian
algorithm cannot be used in the PSMC to model the initial state
radiation (ISR), because of its extremely poor efficiency.
In the latter part of this work we formulate a {\em challenge}
of constructing non-Markovian type of MC algorithm
for ISR PSMC, in which the exact evolution
of the parton distributions is a built-in feature,
like in the Markovian MC case%
\footnote{
  This is contrary to most of the existing MC PSMCs, in which 
  evolved parton distributions come from
  a table obtained using an independent non-MC program.}.
We claim that with present CPU power and sophistication
of the MC techniques such a scenario may be feasible.
With the advent of the future high quality and high statistics
experimental data coming from LHC and HERA, such a solution
is worth to consider.

The layout of the paper is the following: In Section 2 we summarize
basic concepts on the Markovian process and its application
to solution of the QCD evolution equations.
In Section 3 we present exact high precision Markovian MC solution
of the QCD evolution equations.
Section 4 discusses the results and perspectives.

\section{The formalism}
Let us briefly review basic definitions and concepts concerning
the {\em evolution equations} and the {\em Markovian process}.
The generic evolution equations read
\begin{equation}
\label{eq:evolgen}
  \partial_t D_I(t) = \sum_K \Peu_{IK}(t) D_K(t),
\end{equation}
where $t=\ln Q$ and $I$ and $K$ are discrete indexes numbering
states of a certain physical system.
The extension to continuous sets of states is straightforward, see below.
The necessary and sufficient condition for this equation
to have its representation as a probabilistic Markovian process
in the evolution time $t$,
with a properly normalized probability of a forward step,
is the following:
\begin{equation}
\label{eq:condition1}
  \Peu_{II}(t) = -\sum_{K\neq I} \Peu_{KI}(t) \equiv -R_I(t),
  \qquad 
  \Peu_{IK}(t)\geq 0,\;\; \hbox{for}\;\; I\neq K,
\end{equation}
where $R_I(t)\geq 0$ is the {\em decay rate} of the $I$-th parton.
The evolution equation
\begin{equation}
  \partial_t D_I(t) + R_I(t) D_I(t)= \sum_{K\neq I} \Peu_{IK}(t) D_K(t),
\end{equation}
can be easily brought to a homogeneous form
\begin{equation}
\begin{split}
&e^{-\Phi_I(t,t_0)}
     \partial_t \Big( e^{\Phi_I(t,t_0)} D_I(t) \Big)
    = \sum_{K\neq I} \Peu_{IK}(t) D_K(t),\\
&\Phi_I(t,t_0) \equiv \int_{t_0}^{t} dt_1\; R_I(t_1),
\end{split}
\end{equation}
which can be turned into an integral equation
\begin{equation}
\begin{split}
D_I(t)&=e^{-\Phi_I(t,t_0)} D_I(t_0)
  +\int_{t_0}^{t} dt_1\;
   e^{-\Phi_I(t,t_1)}
   \sum_{K} \Peu'_{IK}(t_1)\; D_K(t_1)
\\
\Peu'_{KJ} &\equiv \Bigg\{
           \begin{matrix} \Peu_{KJ}, &  \hbox{for~~~} K\neq J, \\
                          0,         &  \hbox{for~~~} K= J, \\
           \end{matrix}
\end{split}
\end{equation}
and finally can be solved by means of multiple iteration:
\begin{equation}
  \label{eq:IterBasic}
  \begin{split}
  D_K(t) = e^{-\Phi_K(t,t_0)} D_K(t_0)
  +&\sum_{n=1}^\infty \;
   \sum_{K_0...K_{n-1}}
      \prod_{i=1}^n \bigg[ \int_{t_0}^t dt_i\; \Theta(t_i-t_{i-1}) \bigg]
\\ &\times
      e^{-\Phi_K(t,t_n)}
      \prod_{i=1}^n 
          \bigg[ \Peu'_{K_iK_{i-1}}(t_i) 
                 e^{-\Phi_{K_{i-1}}(t_i,t_{i-1})} \bigg]
      D_{K_0}(t_0),
  \end{split}
\end{equation}
where $K\equiv K_n$ is understood, for the brevity of the notation.
The above series of integrals with positively defined integrands
can be interpreted in terms of a random Markovian
process starting at $t_0$ and continuing until $t$
with the following, properly normalized, transition probability
of a single Markovian random step forward,
$(t_{i-1},K_{i-1})\to (t_i,K_i)$,
$t_i> t_{i-1}$, $K_i\neq K_{i-1}$:
\begin{equation}
\label{eq:step}
  \begin{split}  
    &\omega(t_i,K_i| t_{i-1},K_{i-1})
    \equiv  \Theta(t_i-t_{i-1})\;
    \Peu'_{K_iK_{i-1}}(t_i)\;
    e^{-\Phi_{K_{i-1}}(t_i,t_{i-1})},\\
    &\int_{t_{i-1}}^\infty dt_i\; \sum_{K_i}
    \omega(t_i,K_i| t_{i-1},K_{i-1})= 1.
  \end{split}
\end{equation}
Each term in the sum in eq.~(\ref{eq:IterBasic}) can be expressed
as a product of the single step probabilities of
eq.~(\ref{eq:step}) as follows:
\begin{equation}
  \label{eq:Iter2}
  \begin{split}
  D_K(t) =& e^{-\Phi_K(t,t_0)} D_K(t_0)\\
  &+\sum_{n=1}^\infty \;
   \sum_{K_0...K_{n-1}}
      \prod_{j=1}^n \int_{t_0}^t dt_j\;\;
      e^{-\Phi_K(t,t_n)}\;
      \prod_{i=1}^n \omega(t_i,K_i| t_{i-1},K_{i-1})\;
      D_{K_0}(t_0).
  \end{split}
\end{equation}
The Markovian process does not run forever --
it is stopped by the so called
{\em stopping rule}, which in our case is $t_{n+1}>t$.
The variable $t_{n+1}$ is added in eq.~(\ref{eq:Iter2})
by means of inserting an extra integration with the help of an identity
\begin{equation}
  e^{-\Phi_{K_{n}}(t,t_{n})}=
  \int\limits_{t}^\infty dt_{n+1}\; \sum_{K_{n+1}} \omega( t_{n+1},K_{n+1}| t_n,K_n).
\end{equation}
The resulting solution
\begin{equation}
  \label{eq:Markov}
  \begin{split}
  D_K(t) &=
  \int\limits_{t_1>t}dt_{1}\; \sum_{K_{1}} \omega( t_{1},K_{1}| t_0,K_0)
   D_K(t_0)\\
  +&\sum_{n=1}^\infty \;
   \sum_{K_0...K_{n-1}}\;\;
   \int\limits_{t_{n+1}>t} dt_{n+1}\; \sum_{K_{n+1}} \omega( t_{n+1},K_{n+1}| t_n,K_n)
   \prod_{j=1}^n \int\limits_{t_j<t} dt_j\;
\\ &\times
      \prod_{i=1}^n \omega(t_i,K_i| t_{i-1},K_{i-1})\;
      D_{K_0}(t_0).
  \end{split}
\end{equation}
represents clearly a Markovian process, in which,
starting from a point $(t_0,K_0)$ distributed according to
$D_{K_0}(t_0)$, we generate step by step the points $(t_i,K_i), i=1,2,\dots, n+1$,
using transition probability $\omega$, until the stopping rule
$t_{n+1}>t$ acts. 
In such a case the point $(t_{n+1},K_{n+1})$ is {\em trashed}
and the previous point $(t_{n},K_{n})$ is {\em kept} as the last one.
Eq.~(\ref{eq:Markov}) tells us that the points $(t_{n},K_{n})$
obtained in the above Markovian process
are distributed  {\em exactly} according to the solution 
of the original evolution equations%
\footnote{%
The above MC solution
is exact and unbiased contrary to other non-MC methods
which have inherent biases due to choice of finite grid
decomposition into polynomials and other technical artifacts present there.
The main disadvantage of the MC method is its slowness.}
of eq.~(\ref{eq:evolgen}).

Two technical point have to be clarified before we apply the above
solution to the QCD parton distribution evolution equations:
one concerns the $t$ dependence of the $\Peu$ matrix
and another one concerns extension of the above formalism
to continuous indexes in the $\Peu$ matrix.

The transition matrix $\Peu$, in the case of the QCD parton distributions,
includes the running coupling constant $\alpha(t)$
\begin{equation}
\label{eq:alpha}
\Peu(t)_{KL} = \alpha(t)\; P_{KL},
\end{equation}
which depends rather strongly on $t$ at low $t=\ln Q$.
In the above we have explicitly assumed that the {\em kernel} $P_{KL}$
is independent of $t$, which is true in the leading-logarithmic (LL)
case and is not true in the next-to-leading-logarithmic (NLL) case.
However, even in the NLL case this is a very good approximation,
which can be easily corrected in the MC calculation using
an extra weight (and event rejection).
So, we can safely assume the validity of eq.~(\ref{eq:alpha})
without any loss of generality, even beyond the LL level.
If the above factorization (\ref{eq:alpha})
is true, then we may employ the standard trick
which eliminates $t$-dependence from $\Peu$ completely.
This is realized by introducing a new evolution time variable
\begin{equation}
\label{eq:tau}
\tau \equiv \frac{1}{\alpha(t_0)} \int_{t_0}^{t} dt_1\; \alpha(t_1),\quad
\frac{\partial t}{\partial\tau}= \frac{\alpha(t_0)}{\alpha(t)}.
\end{equation}
With this choice we have
\begin{equation}
\label{eq:evolgen2}
  \partial_\tau D_I(\tau) = \sum_K \bar\Peu_{IK} D_K(\tau),\quad
  \bar\Peu_{IK} = \frac{\alpha(t_0)}{\alpha(t)} \Peu_{IK}(t)
   = \alpha(t_0)P_{KL},
\end{equation}
where $\bar\Peu$ does not depend on the new evolution time variable
$\tau$ anymore,
because $\alpha(t)$ in $\bar\Peu$ is effectively replaced by $\alpha(t_0)$.
The rest of the formalism following eq.~(\ref{eq:evolgen}) is the same,
provided we substitute $t\to \tau$  and $\Peu\to\bar\Peu$ everywhere.

Concerning the continuous indexes, indeed in the QCD case we deal
with the composite index $K=\{k,x\}$, where $k$ is the parton type;
let it be $k=G,q,\bar{q}$ in the simple case of gluon and one type of quark,
and $x$, which is a fraction of the energy of the primary (proton)
beam particle carried by the parton -- the usual definition.
The standard parton distribution evolution equations read
\begin{equation}
\label{eq:QCDevol}
\partial_\tau x_1D_{k_1}(\tau,x_1) 
         = \sum_{k_0} 
           \int\limits_{x_1}^{1}\; d x_0\;\;
           \frac{\alpha_S(t_0)}{\pi}\;
           \frac{x_1}{x^2_0} P_{k_1k_0}\Big( \frac{x_1}{x_0}\Big)\;
           x_0 D_{k_0}(\tau,x_0) 
\end{equation}
from which we deduce the following substitution rule
\begin{equation}
 \sum_{K_i} \to \sum_{k_i} \int d x_i,
\end{equation}
while for the matrix elements,
excluding IR region $x_i=x_{i-1}$ from the consideration,
we have the explicit expression
\begin{equation}
 \bar\Peu_{K_iK_{i-1}} = 
     \frac{\alpha_S(t_0)}{\pi}\;
     \frac{x_i}{x^2_{i-1}} P_{k_ik_{i-1}}\Big( \frac{x_i}{x_{i-1}}\Big)\;
     \Theta\Big( 1-\frac{x_i}{x_{i-1}}\Big),
\end{equation}
where $K_i=\{k_i,x_i\}$.

Another two points require now clarification:
the meaning of the {\em diagonal} part $K=J$ in the $\bar\Peu_{K,J}$,
the {\em off-diagonal} part $\bar\Peu'_{KJ}$
and the validity of the condition of eq.~(\ref{eq:condition1})
necessary for the feasibility of the Markovianization.

The problem of defining off-diagonal elements in $\bar\Peu'_{K,J}$
coincides with the problem of defining an infrared (IR) cut-off
and is dealt with in a standard way by introducing a dimensionless
small parameter $\varepsilon\to 0$, which defines/separates
real gluon emission by requiring $x_i/x_{i-1}<1-\varepsilon$.
The off-diagonal (real emission) part
is defined in the usual way%
\footnote{ The ${\cal O}(\varepsilon)$ term is related to the
  fact that for the sake of simplicity we apply 
  the $\Theta(x_j/x_{i}-1+\varepsilon)$
  factor also to the $k_i \neq k_j$ part, which is not IR singular.}
\begin{equation}
\label{eq:BarPeu}
\begin{split}
\bar\Peu'_{K_jK_{i}} 
&=\Theta\Big( 1-\varepsilon-\frac{x_j}{x_{i}} \Big)
     \frac{\alpha_S(t_0)}{\pi}\;
     \frac{x_j}{x^2_{i}} 
     P_{k_jk_{i}}\Big( \frac{x_j}{x_{i}}\Big)\;
  +{\cal O}(\varepsilon)(1-\delta_{k_ik_j}).
\end{split}
\end{equation}
The condition of eq.~(\ref{eq:condition1}),
$\sum_{K_j} \bar\Peu_{K_jK_{i}}=0$,
coincides in fact with the energy conservation sum rule, which reads as follows
\begin{equation}
\label{eq:Rate}
\bar R_{K_i}
= \sum_{K_j} \bar\Peu'_{K_jK_i} 
= \frac{\alpha_S(t_0)}{\pi}
   \sum_{k_j} \int dz_{ji}\;
   z_{ji} P_{k_jk_{i}}( z_{ji})\;
   \Theta( 1-\varepsilon-z_{ji}),
\end{equation}
where $z_{ji}=x_j/x_i$,
and the nice feature is that $\bar R_{K_i}$ depends
only on the parton type $k_i$ and not on $x_i$
(unless we introduce more complicated IR cut).
At this point it is understandable why eq.~(\ref{eq:QCDevol})
was written for the energy distributions $xD(x)$ and not for
the parton distributions $D_K(x)$ themselves.
In the latter case the condition of eq.~(\ref{eq:condition1})
is not generally valid%
\footnote{It is, however, valid for the  non-singlet component.}
and the Markovian MC calculation is not possible.
The significance of the energy sum rules in the evolution equations was
underlined in many works, see for instance 
refs.~\cite{Wosiek:1979ut,Cvitanovic:1980ru}.

\begin{figure}[!ht]
  \centering
  {\epsfig{file=cCanv3.eps.1000x300.singlet_proton,width=120mm}}
  \caption{\sf
    The upper plot shows gluon distribution
    $xD_G(x,Q_i)$ evolved from $Q_0=1$GeV to $Q_i=10,100,1000$GeV
    obtained from {\tt QCDnum16} and  {\tt EvolMC1}, while
    the lower plot shows their ratio.
    The horizontal axis is $\log_{10}(x)$.
    See the text for details.
    }
  \label{fig:G_proton1000x300}
\end{figure}
\begin{figure}[!ht]
  \centering
  {\epsfig{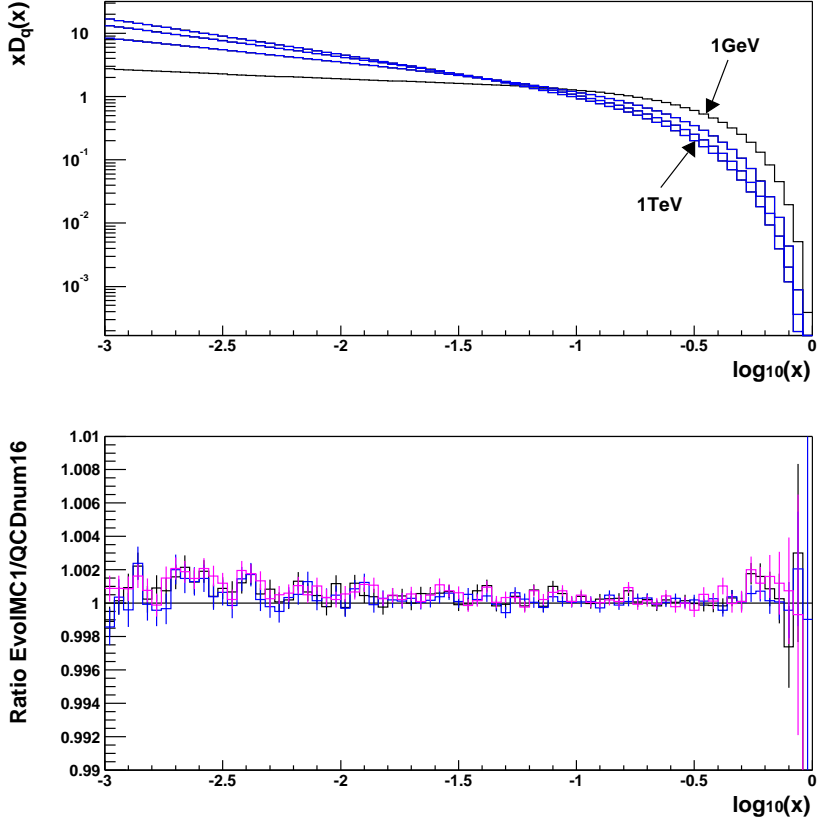}}
  \caption{\sf
    The upper plot shows the singlet quark distribution
    $D_{\bar{q}+q}(x)$ evolved from $Q_0=1$GeV to $Q_i=10,100,1000$GeV
    obtained from {\tt QCDnum16} and  {\tt EvolMC1}, while
    the lower plot shows their ratio.
    The horizontal axis is $\log_{10}(x)$.
    See the text for details.
    }
  \label{fig:Qs_proton1000x300}
\end{figure}

Collecting all the above material we can write the normalized
transition probability for the actual MC Markovian 
calculation as follows
\begin{equation}
\omega(\tau_i,k_i,x_i|\tau_{i-1},k_{i-1},x_{i-1})
=\Theta(\tau_i-\tau_{i-1})\;
  e^{-(\tau_i-\tau_{i-1}) \bar R_{i-1}}\;\;
  \bar\Peu'_{K_iK_{i-1}},
\end{equation}
where the necessary ingredients are provided by 
eqs.~(\ref{eq:BarPeu}) and (\ref{eq:Rate}).
The LL QCD evolution kernels can be found in any QCD textbook
and we skip details of the actual formulas entering the above
equation.

\section{Numerical results}

In the following we present numerical results obtained
for the distributions of three light quarks and gluons
evolved from $Q=1$GeV up to 1TeV in the range $x\in(10^{-4},1)$.
The starting parton distribution at $Q_0=1$GeV is the
proton distribution defined in a conventional way:
\begin{equation}
  \begin{split}
    xD_G(x)        &= 1.9083594473\cdot x^{-0.2}(1-x)^{5.0},\\
    xD_q(x)        &= 0.5\cdot xD_{\rm sea}(x) +xD_{2u}(x),\\
    xD_{\bar q}(x) &= 0.5\cdot xD_{\rm sea}(x) +xD_{d}(x),\\
    xD_{\rm sea}(x)&= 0.6733449216\cdot x^{-0.2}(1-x)^{7.0},\\
    xD_{2u}(x)     &= 2.1875000000\cdot x^{ 0.5}(1-x)^{3.0},\\    
    xD_{d}(x)      &= 1.2304687500\cdot x^{ 0.5}(1-x)^{4.0}.   
  \end{split}
\end{equation}
We use in this exercise the LL version of the evolution kernels
\cite{DGLAP,stirling-book}.
Extension to the NLL level requires to use the NLL kernels
\cite{furmanski80}.
The rest of the MC algorithm/program is the same as in the LL case.
Such an extension will be presented elsewhere.

In figs.~\ref{fig:G_proton1000x300}-\ref{fig:Qs_proton1000x300}
we show numerical results for the gluon and quark distributions evolved
from $Q=1$GeV to higher scales $Q=10, 100, 1000$GeV,
obtained from our Markovian Monte Carlo {\tt EvolMC1}%
\footnote{The actual MC program is rather compact, 
  it consists of about 3k lines of the C++ code.}.
In the plots we also show the results from the standard
program {\tt QCDnum16} evolving parton distributions using
a finite grid of points in the $x$ and $Q$ space.
As we see, these two calculations
agree within 0.2\% for gluons and even better for the quark singlet.
This is clearly  seen from the plotted ratio of the MC result and
the result of {\tt QCDnum16}.
The biggest discrepancy is in the region close to $x=1$,
where all parton distributions are extremely small.
The origin of 0.2\% discrepancy for gluons is not identified unambiguously.
Variation of the grid parameters change the results of {\tt QCDnum16}
quite substantially,
while the variation of the IR cut $\varepsilon$ and of the minimum $x$
parameters in the MC run has no visible influence on  the MC results.
We tend to conclude that the difference between the MC and {\tt QCDnum16}
is due to a numerical bias in the {\tt QCDnum16} program,
although more work is required to reach a firm conclusion.
In the MC we generated $1.2\cdot 10^{9}$ events (96nh CERN units).
It would be possible to increase MC statistics by a factor of 10,
obtaining a sub per-mill statistical errors in the MC results,
if necessary. 

\section{Discussion}
\begin{figure}[!ht]
  \centering
  {\epsfig{file=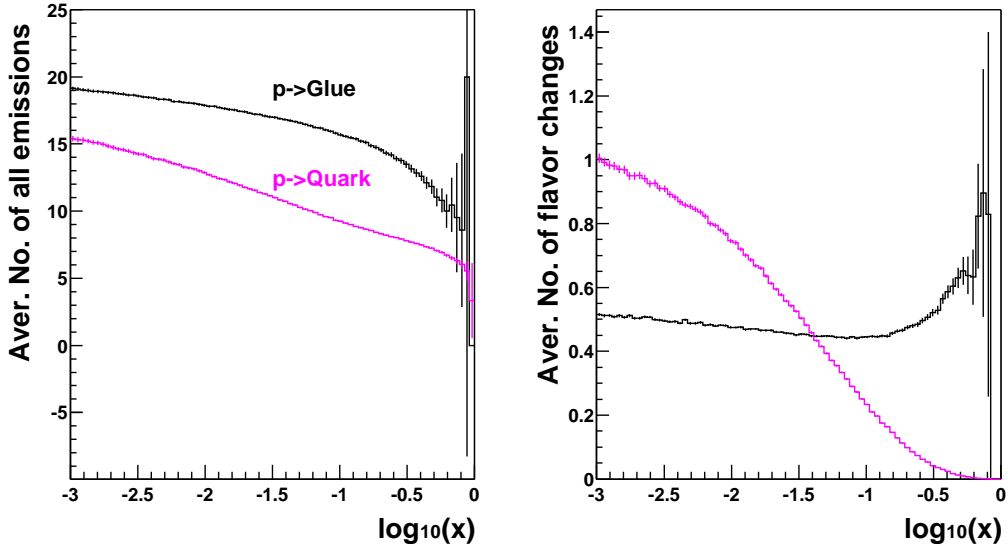,width=140mm}}
  \caption{\sf
    Average multiplicities of all emissions (left plot)
    and of the transitions $G\to q$ and $q\to G$ only (right plot)
    as a function of $\log_{10}(x)$.
    Results are for the proton ``showering'' up to 1TeV scale according
    to Markovian MC with the IR cut $\varepsilon=10^{-4}$.
    }
  \label{fig:multiplicity}
\end{figure}

The main lesson from the above numerical exercises is that with the
modern computing power the
precision of $10^{-4}$ in the solutions of the QCD evolution equations
is within the reach of the MC method
for most of the $x$-range.
MC will always be slower than the non-MC methods, however, it has
no biases related to the finite grid, the use of quadratures,
decomposition into finite series of polynomials,
accumulation of the rounding errors, etc.
MC method can therefore be used to test any other non-MC tool for
the QCD evolution.
Let us stress that
in the MC method there is also no need to split PDFs into a singlet
component and several types of non-singlet components, each of them
evolved separately and combined in the very end of the calculation.
This simplifies the calculation, and the MC Markovian method is clearly
well suited for the evolution of PDFs with many components, for example
QCD+QED, SUSY theory or QCD+electroweak theory at very large energies,
modeling showers of extremely high energetic cosmic rays 
in the intergalactic space (see ref.~\cite{Toldra:2002sn}), etc. 

Let us now elaborate more on the issues related
to the construction of the efficient PSMC for the QCD initial state radiation. 
As it is well known,
the cross section of the hard scattering, especially in the presence
of narrow resonances, discriminates very strongly on
the parton type and $x$ of the parton exiting PSMC
and entering the hard scattering.
The Markovian PSMC does not provide for any control (tagging) of the flavor
and $x$ of the exiting parton, consequently, the overall efficiency
of such a MC for the initial state would be drastically bad.
The solution adopted in the 
most of the existing ISR QCD PSMCs is the so-called backward parton
shower of ref.~\cite{sjostrand85},
in which one starts with the parton next to the hard process,
defining its type and energy $x$, and then the Markovian random walk
continues toward the lower $Q$ and higher $x$,
until the hadron mass scale is reached.
This beautiful algorithm requires, however, the {\em a priori} knowledge
of the parton distribution functions (PDF) at all intermediate $Q$ scales.
These PDFs are, therefore,
tabulated using results of the non-MC evolution programs,
and are also fitted to the deep inelastic scattering (DIS) data.
In other words, the evolution of PDFs is not a built-in feature of 
the standard ISR PSMC.
The two most important reasons for that are the following: 
(i) one is purely technical -- the low efficiency of the standard
Markovian ISR PSMC and 
another one is (ii) the convenience in porting PDFs from DIS data
to hadron-hadron scattering.
We notice that both of these reasons seem nowadays to fade away.
For (i), the CPU power of the present computer systems is 
bigger by the factor $\sim10^4$
than 15 years ago, when the problem of
the efficiency of the PSMC for ISR was considered for the first time.
That widens considerably the range of the practically realizable MC algorithms.
In addition, there exist now working examples of the 
efficient non-Markovian algorithms with
built-in evolution (albeit in the soft approximation) for the pure 
multi-bremsstrahlung process
in the abelian case (QED), see refs.~\cite{Jadach:1987ii,Jadach:2000ir},
which can be treated as a partial solution of the problem
of tagging $x$.
For (ii), PDFs in the future LHC experiments will be constrained
not only by DIS but also by the LHC data ($W$ and $Z$ production) themselves.
One should, therefore, consider the 
genuine MC solution for the QCD evolution as an integral part of the PSMC,
with the aim of fitting the LHC and DIS data simultaneously.
The obvious advantage of such a scenario is that the detector effects
can be removed from the data more efficiently.
No doubt, to realize it in practice will be technically rather challenging.
Nevertheless, we claim that it is feasible and at present
it makes sense to consider it seriously.

In view of the above discussion let us formulate {\em the challenge}
in the construction of the MC event generators:
It would be desirable to construct a MC program/algorithm, which evolves
the parton distribution functions using the QCD NLL $\overline{MS}$ kernels
from the PDFs at low $Q_{\min}$ to higher $Q_{\max}$,
in which we are able to fix beforehand
the type (flavour) and the energy fraction $x$ of the parton
at the scale $Q_{\max}$.
The {\em a priori} knowledge of the PDFs at all intermediate $Q$s
should not be required/used; the input from the perturbative QCD
would be the expressions for the kernels. 
The PDFs at all intermediate $Q$s would be the result from the MC.

Additional remarks:
(i) We do not assume that the solution is based on the Markovian process.
Any, even brute force solution, but feasible with the present
CPU power, is worth consideration. 
(ii) The above challenge concerns primarily the $(Q,x)$
space. The construction of a full scale 
parton shower MC on top of the above two-dimensional MC we treat 
as a separate (important) issue.
(iii) In our opinion, solving the above technical challenge would 
open new interesting avenues in the realm of the MC event generators
for DIS and hadron-hadron scattering.

For the moment we do not offer any solution of our own.
We can see one hint, which comes from the exercises we have done using the
presented MC of the Markovian type.
In the left hand side plot in Fig.~\ref{fig:multiplicity} 
we show the average number of all ``emissions'' in the 
Markovian process on the way from $Q=$1GeV to $Q=$1TeV,
for the proton case, as a function of the final $x$ at $Q=$1TeV.
The IR cut-off $\varepsilon=10^{-4}$ is used.
Due to a sizable value of $\alpha_S$ and large phase space
the average number of emissions is about 20 (for a single proton beam).
However, the average number of the transitions 
$q,\bar{q}\to G$ and $G\to q,\bar{q}$,
shown in the right hand side plot in the same figure,
is much smaller -- it is in the range from 0.5 to 1.
This feature suggests the brute force solutions in which 
$q,\bar{q}\to G$ and $G\to q,\bar{q}$ transition are pretabulated
(solving the problem of tagging the parton type),
while the pure bremsstrahlung transition $G\to G$ and $q\to q$
are treated using the algorithm of the QED
(with the constrained final $x$).
However, we hope that more elegant and efficient solution can be found.
Finally, we would like to refer the reader to extensive literature on the subject
of the QCD evolution and parton shower MC algorithms,
which can be found for instance in refs.~\cite{stirling-book,khoze-book}.

\vspace{10mm}
\noindent
{\bf Acknowledgments}\\
We would like to thank W. P\l{}aczek for the useful discussions.


\begingroup\begin{thebibliography}{10}

\bibitem{DGLAP}
L.N. Lipatov, Sov. J. Nucl. Phys. {\bf 20} (1975) 95;\\ V.N. Gribov and L.N.
  Lipatov, Sov. J. Nucl. Phys. {\bf 15} (1972) 438;\\ G. Altarelli and G.
  Parisi, Nucl. Phys.{\bf 126} (1977) 298;\\ Yu. L. Dokshitzer, Sov. Phys. JETP
  {\bf 46} (1977) 64.

\bibitem{webber88}
G.~Marchesini and B.~Webber, {\em Nucl. Phys.} {\bf B310} (1988) 461.

\bibitem{qcdnum16}
M.~Botje, ZEUS Note 97-066, http://www.nikhef.nl/~h24/qcdcode/.

\bibitem{blumleinn96}
J.~Blumlein {\em et al.} in {\em Future Physics at \uppercase{HERA}}
  (G.~Ingelman, A.~De~Roeck, and K.~R, eds.), vol.~1, p.~23, 1996.

\bibitem{Wosiek:1979ut}
J.~Wosiek and K.~Zalewski, {\em Acta Phys. Polon.} {\bf B10} (1979)
667--669.

\bibitem{Cvitanovic:1980ru}
P.~Cvitanovic, P.~Hoyer, and K.~Zalewski, {\em Nucl. Phys.} {\bf B176} (1980)
429.

\bibitem{stirling-book}
R.~K.~Ellis, W.~Stirling, and B.~Webber, {\em QCD and Collider Physics}.
\newblock Cambridge University Press, 1996.

\bibitem{furmanski80}
G.~Curci, W.~Furmanski, and R.~Petronzio, {\em Nucl. Phys.} (1980) 27.

\bibitem{Toldra:2002sn}
R.~Toldra,
\href{http://www.arXiv.org/abs/astro-ph/0201151}{{\tt astro-ph/0201151}}.

\bibitem{sjostrand85}
T.~Sjostrand, {\em Phys. Lett.} (1985) 321.

\bibitem{Jadach:1987ii}
S.~Jadach, MPI-PAE/PTh 6/87.

\bibitem{Jadach:2000ir}
S.~Jadach, B.~F.~L. Ward, and Z.~Was, {\em Phys. Rev.} {\bf D63} (2001) 113009,
\href{http://arXiv.org/abs/hep-ph/0006359}{{\tt hep-ph/0006359}}.

\bibitem{khoze-book}
Y.~Dokshitzer, V.~Khoze, A.~Mueller, and S.~Troyan, {\em Basics of Perturbative
  QCD}.
\newblock Editions Frontieres, 1991.

\end{thebibliography}\endgroup

\providecommand{\href}[2]{#2}

\end{document}